# Model-based clustering for identifying disease-associated SNPs in case-control genome-wide association studies

Yan Xu[1,¶], Li Xing[1,¶], Jessica Su[2], Xuekui Zhang[1,*], Weiliang Qiu[2]

[1] Department of Mathematics and Statistics, University of Victoria, Victoria, BC, Canada

[2] Channing Division of Network Medicine, Brigham and Women's Hosptial/Harvard Medical School, Boston, MA, USA

\* Corresponding author

E-mail: Xuekui@UVic.ca

¶These authors contributed equally to this work.

# Abstract

Genome-wide association studies (GWASs) aim to detect genetic risk factors for complex human diseases by identifying disease-associated single-nucleotide polymorphisms (SNPs). The traditional SNP-wise approach along with multiple testing adjustment is over-conservative and lack of power in many GWASs. In this article, we proposed a model-based clustering method that transforms the challenging high-dimension-small-sample-size problem to low-dimension-large-sample-size problem and borrows information across SNPs by grouping SNPs into three clusters. We pre-specify the patterns of clusters by minor allele frequencies of SNPs between cases and controls, and enforce the patterns with prior distributions. In the simulation studies our proposed novel model outperform traditional SNP-wise approach by showing better controls of false discovery rate (FDR) and higher sensitivity. We re-analyzed two real studies to identifying SNPs associated with severe bortezomib-induced peripheral neuropathy (BiPN) in patients with multiple myeloma (MM). The original analysis in the literature failed to identify SNPs after FDR adjustment. Our proposed method not only detected the reported SNPs after FDR adjustment but also discovered a novel BiPN-associated SNP rs4351714 that has been reported to be related to MM in another study.

# Introduction

Genome-wide association studies (GWASs) aim to detect genetic risk factors for complex human diseases by identifying disease-associated single-nucleotide polymorphisms (SNPs). The most commonly-used approach in GWASs is the SNP-wise approach, in which a test of association is performed for each SNP, and then the P-values are adjusted for multiple testing. However, because of multiple testing adjustment to a huge number (> 1 million) of tests in GWAS, this approach often lacks power. Multiple testing adjustment uses no information other than P-values, which insufficiently models the relationships among SNPs, and need to be improved.

To reduce the number of tests, SNP-set analysis has been proposed (e.g., Wu et al. 2010[1]; Dai et al. 2013[2]; Lu et al. 2015[3]; Cologne et al. 2018[4]). The idea is to use SNP sets to replace individual SNPs. Hence, the number of tests can be reduced. However, it is challenging to define SNP sets. One approach is to define SNP sets based on existing biological knowledge. However, biological knowledge is subject to error. Poor quality of SNP-set can lead to low power (Fridley and Biernacka, 2011[5]).

Penalized regression approach has also been proposed in GWASs. For instance, linear mixed models (e.g., Kang et al. 2010[6]; Lippert et al. 2011[7]; Zhou and Stephens 2012[8]) treat the effect of the SNP marker of interest as fixed, with the effects of all other SNP markers as normally distributed random effects. This process is repeated in turn for every SNP marker. However, it is a paradox to treat markers as fixed for inference but then otherwise as random to account for population structure for inference on association with other markers (Goddard et al. 2016[9]; Chen et al. 2017[10]). Bayesian hierarchical regression models treat the effects of all SNPs as random effects with either local priors or non-local priors (Mallick and Yi 2013[11]; Fernando and Garrick 2013[12]; Wang et al. 2016[13]; Chen et al. 2017[10]). Noticing that penalized regression methods often lead to large number of false positives and Bayesian regression methods are computationally very expensive, Sanyal et al. (2018)[14] proposed a non-local prior based iterative SNP selection tool for GWASs, which enables borrowing information across SNPs and can utilize the dependence structure across SNPs. However, all of these methods are for quantitative outcomes (e.g., height) in GWASs.

Several methods have been proposed to increase statistical power based on borrowing information across gene probes via mixture models. For example, Gamma-Gamma model (GG)[15], Log-Normal-Normal (LNN)[16], extended GG (eGG)[17], extended LNN (eLNN)[17], eLNN for paired data[18], and Marginal Mixture Distributions (GeneSelectMMD)[19] have been proposed for gene microarray data, and edgeR[20], DESeq[21, 22], and DESeq2 [23] have been proposed for next-generation sequencing (RNAseq) data. All these methods have been successfully applied to either gene microarray data analysis (continuous-scale data) or RNAseq data analysis (count data). However, to the best of our knowledge, no methods have been proposed to borrow information across SNPs (categorical variables with three levels of genotype) to analyze case-control GWAS data that have binary phenotype (cases vs. controls).

In this article, we proposed a novel model-based clustering method for case-control GWASs (binary phenotype) by transforming the challenging high-dimension-small-sample-size problem to low-dimension-large-sample-size problem. Specifically, we treat SNPs as "samples" and subjects as "variables" and aim to cluster SNPs to three groups: (1) SNPs with minor allele frequencies (MAFs) higher in cases than in controls; (2) SNPs with MAFs lower in cases than in controls; and (3) SNPs with MAFs in cases same as in controls. For a given SNP, we assume its genotypes follow a multinomial distribution and the MAF follows a beta distribution. We also assume that the cluster proportions follow a Dirichlet distribution. In our method, we pre-specify the patterns of clusters by minor allele frequencies (MAFs) of SNPs between cases and controls, and enforce the patterns with the guide of prior distributions. The proposed model-based clustering method can improve the power of detecting disease-associated SNPs by borrowing information across SNPs. For details, please refer to the METHODS Section.

Our method is motivated by our investigation of the genetic risk factors of the bortezomib-induced peripheral neuropathy (BiPN) in treating Multiple Myeloma (MM) by using GWASs. MM is a type of cancer that causes a group of plasma cells (a type of white blood cell in the bone marrow that helps fight infections by making antibodies that recognize and attack germs) to be cancerous[24]. The MM cancer cells produce abnormal proteins that can cause complications that can damage the bones, the immune system, kidneys, and red blood cell. MM is the third most common blood cancer in the United States. Bortezomib is a first-in-class proteasome inhibitor to treat MM[25, 26]. However, Bortezomib has some side effects, such as the development of a painful, sensory peripheral neuropathy (PN)[27-29]. Bortezomib could induce neurotoxicity in neuronal cells by several mechanisms that lead to apoptosis[30]. The symptoms of the BiPN include neuropathic pain and a length-dependent distal sensory neuropathy with a suppression of reflexes. Due to BiPN, patients often discontinue bortezomib treatment despite a good response to the therapy[31]. If we could identify patients at the risk of developing BiPN, physicians then can choose alternative therapies, such as using weekly, reduced-dose, or subcutaneous approaches. However, BiPN mechanisms are mostly unknown. It has been shown that a higher cumulative dose is likely to predict the increase of severity of BiPN[32, 33]. Pre-existing neuropathies, comorbidities (like diabetes mellitus) or myeloma-related peripheral nerve damage may also increase the risk of developing BiPN[34, 35]. Meregalli (2015)[36] provided a review of bortezomib-induced neurotoxicity.

The inter-individual difference in the onset of BiPN indicates that genetics plays an important role. The candidate gene approaches have identified a few single nucleotide polymorphisms (SNPs) associated with the development of BiPN[28, 37-39]. For example, Broyl et al. (2010)[28] identified 20 BiPN-associated SNPs after examining 3,404 candidate SNPs. The sample sizes in Broyl et al.'s (2010) study[28] are relatively small. Seven of the 20 SNPs were identified by comparing 13 grade 2-4 BiPN patients after one cycle of bortezomib treatment with 147 no-BiPN patients (rs2251660, rs4646091, rs1126667, rs434473, rs7823144, rs1879612, and rs1029871). The other 13 SNPs were identified by comparing 49 grade 2-4 BiPN patients after two or three cycles of bortezomib treatment with 80 no-BiPN patients (rs1799800, rs1799801, rs2300697, rs1059293, rs2276583, rs189037, rs10501815, rs664677, rs664982, rs6131, rs1130499, rs4722266, and rs2267668). Corthals et al. (2011)[38] conducted a candidate SNP analysis with larger sample sizes than Broyl et al. (2010)[28] did, which revealed associations with BiPN based on 2,149 SNPs using a discovery set with 238 samples and a validation set with 231 samples. However, after adjusting for multiple testing, no significant SNPs were identified. Favis et al. (2011)[39] conducted several survival analyses based on 2,016 SNPs and identified five BiPN associated SNPs (rs4553808, rs1474642, rs12568757, rs11974610, and rs916758) in the discovery set (139 samples) after adjusting for multiple testing. However, none of these five SNPs were validated in the validation set (212 samples).

GWAS could be used to unbiasedly identify genetic variants that will have a direct or indirect effect on drug sensitivity[29]. NHGRI GWAS Catalog [40, 41](https://www.ebi.ac.uk/gwas/) lists only two GWA studies that have been performed to identify SNPs associated with BiPN for MM patients. The first GWAS was performed by Magrangeas et al. (2016)[29], who identified one BiPN-associated SNP (rs2839629) based on 370,605 SNPs using a discovery set (469 samples) and a validation set (114 samples). However, results did not reach a genome-wide significance level. The second GWAS was conducted by Campo et al. (2018)[42], who identified four BiPN-associated SNPs (rs6552496, rs12521798, rs8060632, and rs17748074) based on 646 samples. Again, the results did not reach a genome-wide significance level. Moreover, each of the four lists of BiPN-associated SNPs listed above (the 20 SNPs identified by Broyl et al. (2010)[28]; the five SNPs identified by Favis et al. (2011)[39]; the one SNP identified by Magrangeas et al. (2016)[29]; and the four SNPs identified by Campo et al. (2018)[42]) was not replicated in the other three studies.

Note that the existing two GWASs[28, 42] used SNP-wise approaches (i.e., performing one association test per SNP), which over-adjusts for multiple tests due to insufficiently modeled relationships among SNPs (i.e., true FDRs/FWERs are smaller than nominal FDR/FWER levels), and hence are not powerful enough when sample sizes are relatively small. Therefore, the missing heritability[43] of BiPN could be due to less powerful statistical methods. Novel statistical methods are needed, they could borrow information across SNPs and better control FDR at nominal levels than the traditional over-conservative multiple testing adjustment approaches.

Our novel method for SNP discovery is a model-based clustering approach. In our method, information can be shared across SNPs by grouping SNPs into three clusters. We pre-specify the patterns of clusters by minor allele frequencies (MAFs) of SNPs between cases and controls, and enforce the patterns with the guide of prior distributions. Using simulation studies and re-analysis of real data from Magrangeas et al. (2016)[29], we demonstrated that our method out-performs traditional approaches. In particular, compared to SNP-wise approaches, our method better controls FDR at a nominal level, and therefore it has a better sensitivity.

## Results

**Results of simulation studies**

We conducted simulation studies to compare the performance of our model-based clustering method with the SNP-wise approach (e.g., logistic regression followed by multiple testing adjustment ). For our method, data analysis of each simulated dataset uses two different ways to choose values of hyper-parameters of the prior distributions for MAFs (obtained from either truncated $\text{Beta}(2,5)$ or empirical distribution via moment matching approach). Details on the design of simulation studies and the choice of hyper-parameters for MAF priors are explained in the 'METHODS' Section.

Figure 1 shows the simulation results for 500,000 SNPs with 200 effective SNPs using four different MAF distributions for data generation and two different sample sizes. The results of comparing sensitivity are presented in the right panel of the figure, where the boxplots represent differences between our method and the SNP-wise approach, which were obtained by considering sensitivity of our method using truncated $\text{Beta}(2,5)$ (colored in light grey) or empirical distribution (colored in dark grey) minus sensitivity of the SNP-wise approach for each simulated dataset. Our method outperformed the SNP-wise approach in sensitivity, since all of the boxes are on the right-hand side of the vertical dashed line 0. Boxplots in the middle panel show FDRs for the SNP-wise approach, our method using truncated $\text{Beta}(2,5)$ for analysis, and our method using empirical estimates for analysis. FDR of detected SNPs using our method is much closer to the nominal level 0.05 (vertical dashed line) than the SNP-wise approach for every setting presented in the figure. We conducted Wilcoxon signed rank tests to compare our method and the SNP-wise approach on |FDR-0.05| and on sensitivity for the settings in Figure 1. All of tests were significant with small P-values ($< 0.0093$; see Supplementary Table S2), confirming that the differences observed from the parallel boxplots are statistically significant. By comparing the top four rows with the bottom four rows in Figure 1, we also observed that the improvement of our method over the traditional approach becomes more prominent when the sample size of the study was smaller.

In all other settings of simulations studies, compared with the traditional approach, our method consistently showed higher sensitivities and FDRs closer to the nominal level of 0.05 (see Figures S1-S5 in Supplementary File S4). Results from Wilcoxon signed rank tests for these settings are also provided in Supplementary Table S2 with all P-values smaller than the significance level 0.05. Even when we incorrectly specified the prior distributions for analysis (e.g., when we used different values between data generation and data analysis for hyper-parameters $\alpha$ and $\beta$), our method still outperformed the traditional SNP-wise approach (see rows 2-4 and 6-8 in Figure 1 and S1-S5 in Supplementary File S4).

In summary, our method performed better in discovering effective SNPs associated with the outcome. The traditional SNP-wise approach over-penalizes multiple testing and insufficiently utilizes the shared information among SNPs. When targeting on the nominal FDR level 0.05, the traditional approach always had a true FDR less than the nominal level, which reduces sensitivity.

**Results of real data analysis**

From the Gene Expression Omnibus (GEO) (https://www.ncbi.nlm.nih.gov/geo/), we downloaded two SNP datasets (GSE65777 and GSE66903) that were used by Magrangeas et al. (2016)[29] to evaluate the genetic effects on developing severe bortezomib-induced peripheral neuropathy (BiPN). The discovery dataset (GSE65777) contains 909,622 SNPs and 469 newly-diagnosed patients with multiple myeloma treated with bortezomib. The goal was to compare SNP genotypes from 155 patients with grade $\geq 2$ BiPN with those from 314 MM patients with grade 1 BiPN or no BiPN. The validation dataset (GSE66903) contains 795,734 SNPs and 116 MM patients treated with bortezomib. The goal of the validation study was to compare 41 bortezomib-treated grade $\geq 2$ BiPN patients with 75 bortezomib-treated control patients. A genome-wide association study of the 370,605 SNPs after quality control (QC) on the discovery data GSE65777 was conducted by Magrangeas et al. (2016)[29]. In our analysis, 247,372 SNPs that passed our QC criteria, which is basically the same as the criteria used by Magrangeas et al. (2016)[29] except for two minor changes (see Section D of Supplementary File S3 for details on our QC steps).

We first re-analyzed discovery data (GSE65777) with SNP-wise approaches. The association between the outcome and each SNP was tested by both logistic regression and the Cochran-Armitage test. No SNP is significant after multiple testing adjustment at FDR level 0.1. Note SNPs reported by Magrangeas et al. (2016)[29] were not detected with the genome-wide threshold $5 \times 10^{-8}$, but with a much larger P-value threshold $10^{-5}$.

We then analyzed the discovery dataset (GSE65777) with our method. The values for hyper-parameters $\alpha$ and $\beta$ were set to their moment estimates from SNP data. Table 1 lists the significant SNPs detected based on the combinations of settings of two different pseudo counts and two targeted FDR levels. Since all significant SNPs detected by our method have been adjusted for FDR, our method is more powerful in detecting SNPs than the traditional approach.

We next analyzed the validation dataset (GSE66903) to validate the significant SNPs in Table 1, using exactly the same approach as Magrangeas et al. (2016)[29]. The SNP rs2839629 can be validated (with a P-value of 0.0324 after multiple adjustment controlling FDR level at 0.1), which is detected in discovery dataset using a pseudo count of (3, 3, 3) and a detection rule of $\widehat{\text{FDR}} < 0.1$ (Table 2).

We tried analyzing the data with a different pseudo count (5, 5, 5), and get exactly the same results as using (3, 3, 3). When we used a stronger prior by increasing the pseudo count to (20, 20, 20), we obtained six SNPs that were assigned to the clusters of significant SNPs. Among the six SNPs, rs4351714 is a possible novel SNP to BiPN, which locates in the intron region of gene *KDM5B* and is proved to be associated with multiple myeloma[44, 45], but no existing literature has reported that rs4351714 is associated to BiPN. *KDM5B* is known as a member of the *KDM5* subfamily that serves as transcriptional co-repressors, specifically catalyzing the removal of all possible methylation states from lysine 4 of histone H3 (H3K4me3/me2/me1). It has been linked to control of cell proliferation, cell differentiation and several cancer types. By employing a *KDM5* enzymes inhibitor in myeloma cells, a higher quartile of *KDM5B* expression was found to be associated with shorter overall survival in myeloma patients[45].

## Discussion

We proposed a novel model-based clustering method to characterize the association between SNPs and a binary outcome in case-control genome-wide association studies. Compared with the traditional SNP-wise approach, it has advantages in efficiently utilizing the data, since we account for the relationships among SNPs in the model. Our novel method has two major advantageous features.

First, compared to the traditional method, our method provides more power to detect true SNPs associated with the outcome and better controls FDR at a nominal level without an over-conservative penalty from multiple testing adjustment. In the traditional SNP-wise approach, an association between the outcome and each SNP is tested separately, and then their P-values are adjusted for multiple testing. The multiple testing adjustment is purely based on P-values, which insufficiently utilizes the relationship among SNPs. In contrast, we group SNPs into clusters according to the pattern of their MAFs, allowing SNPs with similar patterns to share information with each other. The advantage of this feature of our method is demonstrated in both simulation studies and the re-analysis of the real data from a study on patients with multiple myeloma treated with bortezomib.

Second, our model-based clustering method can handle millions of SNPs, which makes it tractable to "simultaneously" model a huge number of SNPs from the ultra-high dimensional GWAS data. Though the model is complex and involves millions of parameters, making the algorithm of model fitting quite challenging to implement, we integrate out the nuisance parameters (i.e., remove nuisance parameters from model likelihood by averaging likelihood over the distribution of nuisance parameters). By only dealing with the essential parameters in the model fitting process, we make the algorithm feasible without losing information.

Note that our novel model-based clustering method is different from the standard clustering methods. Standard clustering methods are an unsupervised learning method that discovers patterns freely from data. In contrast, supervised learning methods train models with both outcomes and predictors. Our method is in between. We specify cluster structures and enforce characteristics of each cluster by model priors, but we do not have true cluster memberships as the outcome to train the model. So we call our method a pseudo-supervised learning approach.

In our pseudo-supervised learning approach, the prior modeling is the key component, which regulates SNPs into the correct clusters. In the machine learning literature, regularized regression models (e.g., Lasso and Elastic Net) always have their Bayesian equivalent counterpart (Bayesian Lasso[46] and Bayesian Elastic Net[47]). In these Bayesian approaches, shrinkage priors are used to achieve the equivalent penalty effect in regularized regressions. We adopt this idea and let the prior distributions guide the discovery of the patterns. In our model, the number of clusters is fixed, and the pattern of each cluster is described and enforced by the prior distributions.

Using the same validation approach as in Magrangeas et al. (2016)[29], we validated the same SNP identified by Magrangeas et al. (2016)[29]. No more SNPs were validated partly due to the inconsistency of signals between the discovery dataset and the validation dataset. First, not many SNPs have strong signals in both studies. See Table (a) in Supplementary File S5 where we ranked SNPs by raw P-value from smallest to largest. Among the top 1000 ranked SNPs in the discovery dataset with smallest raw P-values, only 30 SNPs have a raw P-value <0.05 for the validation data (see Table (b) in Supplementary File S5). Also, the ranks of these 30 SNPs are quite low in the validation set. Second, many SNPs have signals from opposite directions between the discovery set and the validation set. Among the top 1000 SNPs, 513 of them have opposite sign of MAF difference between cases and controls in the two datasets (Table (a) in Supplementary File S5). This means more than half of the top-ranked SNPs are risk factors in one study but protective factors in another study. Among the 30 SNPs with reasonably strong signals in both studies, as mentioned above, nine SNPs showed opposite directions of MAF difference between cases and controls.

## Conclusion

Genome-wide association studies (GWASs) aim to detect genetic risk factors for complex human diseases by identifying disease-associated single-nucleotide polymorphisms (SNPs). We developed a novel method for SNP discovery based on model-based clustering, which can also be considered as a pseudo-supervised machine learning approach. We compared our method with the traditional SNP-wise approach through simulation studies and a real data analysis. The traditional SNP-wise approach is over-conservative since its adjustment for multiple testing is purely based on P-values, insufficiently accounting for the relationship between SNPs. Therefore, its true

FDR is always less than the nominal level and has less power to detect true signals. In comparison, our method can better control FDR at nominal level and detect more effective SNPs. In addition, our method simultaneously models all SNPs but makes computing feasible by integrating out nuisance parameters from the model.

In the re-analysis of the real data from Magrangeas et al. (2016)[29], the traditional method failed to detect any significant SNP after FDR adjustment. In contrast, our proposed method not only detected effective SNPs at the genome-wide significance level, which were reported in Magrangeas et al. (2016)[29] with a much larger P-value threshold than the genome-wide significance level, but also identified a novel BiPN-associated SNP rs4351714 that has been proven to be associated with multiple myeloma.

In summary, our method outperforms the traditional SNP-wise approach in SNP discovering from case-control GWAS.

## Methods

**Notations and 3-cluster mixture models**

Suppose we measure genotypes of G SNPs for $n_x$ MM patients with BiPN (cases) and $n_y$ MM patients without BiPN (controls). Our goal is to identify a subset of SNPs that are significantly associated with the risk of developing BiPN. For each SNP, we code its genotype as: 0 minor allele (wild-type homozygote), 1 minor allele (heterozygote), and 2 minor alleles (mutation homozygote). We assumed Hardy Weinberg Equilibrium (HWE) for each SNP. Then the genotype frequencies of a SNP can be expressed as functions of the Minor Allele Frequency (MAF) θ: Pr(genotype=0) = $(1-\theta)^2$, Pr(genotype=1) =

$2\theta(1-\theta)$, and Pr(genotype=2) =$\theta^2$. If a SNP has significantly different MAFs between cases and controls, then this SNP is associated with the risk of developing BiPN.

We assume that there are three clusters of SNPs: (1) no effect cluster: cluster of SNPs having similar MAF between cases and controls (denoted as cluster 0); (2) positive effect cluster: cluster of SNPs having significantly higher MAF in cases than in controls (denoted as cluster +); and (3) negative effect cluster: cluster of SNPs having significantly lower MAF in cases than in controls (denoted as cluster −). That is, a MM patient with minor alleles of any SNP in cluster + tends to develop BiPN after receiving Bortezomib (i.e., having positive tendency in developing BiPN), while a MM patient with minor alleles of any SNP in cluster – tends to be protective from developing BiPN (i.e., having negative tendency in developing BiPN). SNPs in cluster 0 do not affect developing BiPN.

We use vectors of binary latent variables to describe the unknown cluster memberships of SNPs. For a given SNP g, its cluster membership can be expressed as $\mathbf{z_g} = (z_{g,0}, z_{g,+}, z_{g,-})$. Let $z_{g,k} = 1$ indicate that SNP g belongs to cluster k, and $z_{g,k} = 0$ otherwise, where $k = 0, +, -$ is the index of clusters. Here the indicators must satisfy $\sum_{k\in\{0,+,-\}} z_{g,k} = 1$. Let $\boldsymbol{\pi} = (\pi_0, \pi_+, \pi_-)$, satisfying $\sum_{k\in\{0,+,-\}} \pi_k = 1$, be the proportion of SNPs in three clusters, which is called a mixture proportion. So the cluster membership $z_g$ follows a multinomial distribution as $\Pr(z_g|\boldsymbol{\pi}) = \text{Multinomial}(1, \boldsymbol{\pi}) = \prod_{k\in\{0,+,-\}} \pi_k^{z_{g,k}}$. We assume a Dirichlet prior Dir($\mathbf{b}$) on $\boldsymbol{\pi}$, $\mathbf{b} = (b_0, b_+, b_-)$. Values of $\mathbf{b}$ can be interpreted as pseudo count [48, 49] for each cluster.

The primary objective of the data analysis is to estimate the posterior distribution of $z_g$ given all observed SNP data and values of all model parameters, which will be used for the inference about which SNPs are significantly associated with the outcome.

What we described above is mixture models with three components. Next, we discuss in details about how the mixture models can be derived from mixture of Bayesian hierarchical models by modeling different patterns of the relationship between SNPs and binary outcomes of case-control status inside each cluster.

**Model and prior of genotype and MAFs**

For a given SNP $g$ of patient $i$ under condition $d$, we denote its genotype and minor allele frequency (MAF) in cluster $k$ as $S_{g,d,i}$ and $\theta_{g,d,k}$ respectively, where $d = x$ (case) or $y$ (control), $g = 1, \ldots, G$, $i = 1, \ldots, n_d$, and $k=0, +, -$. We modeled the distribution of the genotype $S_{g,d,i}$ of SNP $g$ by the following multinomial distribution:

$$g(S_{g,d,i}|\theta_{g,d,k}) = \text{Multinomial}\{1, [\theta_{g,d,k}^2, 2\theta_{g,d,k}(1 - \theta_{g,d,k}), (1 - \theta_{g,d,k})^2]\} \quad (1)$$

Note for SNPs in cluster $0$, we have $\theta_{g,x,0} = \theta_{g,y,0}$; For SNPs in cluster $+$, we have $\theta_{g,x,+} > \theta_{g,y,+}$; and for SNPs in cluster $-$, we have $\theta_{g,x,-} < \theta_{g,y,-}$. These conditions will be enforced by prior distributions of MAF discussed below.

Patterns of MAFs in different clusters can be distinguished and enforced by different prior distributions. Shared prior distributions of MAFs allow them to borrow strength from each other. If a SNP has no effect on the outcome, it should have the same MAF in cases and controls. Hence, we use the same conjugate prior for both cases and controls: $\theta_{g,d,0} \sim \text{Beta}(\alpha, \beta)$. We denote its Probability Density Function(PDF) as $h(\cdot)$, and use this PDF to help construct PDFs of the prior distributions for the other two clusters.

For a SNP in cluster $+$, i.e., SNPs having larger MAF in cases than in controls, we define a "half-bell shape" prior which enforces patterns of cluster $+$ with probability 1. We assign a bivariate prior $(\theta_{g,x,+}, \theta_{g,y,+})$ with PDF $2h(\theta_{g,x,+})h(\theta_{g,y,+})I(\theta_{g,x,+} > \theta_{g,y,+})$, where $I(a)$ is the indicator function taking value 1 if a >0, and value 0 otherwise. Note in this PDF, the term $h(\theta_{g,x,+})h(\theta_{g,y,+})$ can be considered as a "bell shape" bivariate distribution of independently and identically distributed (i.i.d.) variables $\theta_{g,x,+}$ and $\theta_{g,y,+}$. The indicator function $I(\theta_{g,x,+} > \theta_{g,y,+})$ enforces the constraint $\theta_{g,x,+} > \theta_{g,y,+}$ with probability 1. It "flattens" half of the bivariate distribution to change the shape into "half-bell", which makes a PDF of $\theta_{g,x,+}$ and $\theta_{g,y,+}$ satisfy our constraints with probability 1. The constant "2" in prior PDF ensures it is a proper PDF (i.e., integrates to 1). Similarly, For a SNP in cluster $-$, we use the other "half-bell" prior by assuming a bivariate prior $(\theta_{g,x,-}, \theta_{g,y,-})$ with PDF $2h(\theta_{g,x,-})h(\theta_{g,y,-})I(\theta_{g,x,-} < \theta_{g,y,-})$.

In GWASs, the set of parameters $\{\theta_{g,d,k}\}$ contains huge numbers of elements due to a large number of SNPs to be considered. Note that all $\{\theta_{g,d,k}\}$ are nuisance parameters, since they are not used for final inference of association between SNPs and outcomes. Hence, we integrate out all nuisance parameters $\{\theta_{g,d,k}\}$ from the model by averaging the complete likelihood over their prior distributions (more details in Section A of Supplementary File S3). Marginalization reduces a huge number of unnecessary parameters from our model likelihood, and thus it makes model fitting feasible and more tractable.

**Choice of values for hyper-parameters**

Our method is robust for different settings of hyper-parameters, as long as the choice of their values is not extremely unreasonable. More details are given in simulation studies.

The hyper-parameters $\alpha$ and $\beta$ are estimated by the moment matching approach based on the distribution of MAFs (detailed formulas are given in Section C of Supplementary File S3). In practice, if GWAS data contain a sufficient amount of SNPs as well as patient samples, we can estimate an empirical distribution of MAFs from all observed SNPs. When the dataset is not rich enough, we recommend using the left-truncated Beta distribution $\text{Beta}(2, 5)$. The distribution $\text{Beta}(2, 5)$ is estimated from the distribution of MAF provided by Keinan et al. (2007)[50]. The truncated range is from the minimum MAF observed in the SNP data after quality control to 0.5. Note that both empirical distribution and the truncated $\text{Beta}(2, 5)$ need to be approximated by a Beta distribution using the moment matching approach, which we call a Beta-approximation. This approximation can greatly speed up computation by using a conjugate prior. The detailed algorithm about calculation of ($\alpha$, $\beta$) in these two situations is given in Section C of Supplementary File S3. Even if parameters are incorrectly specified, our method still can achieve better performance compared to the traditional SNP-wise approach (shown in simulation studies).

Values of the hyper-parameters $(b_0, b_+, b_-)$ in Dirichlet prior can be assigned as small integers, e.g., (3, 3, 3), which is equivalent to a weak prior. Changing it to other small integers (e.g., using (5, 5, 5)) will not affect final results. Using a very strong prior, e.g., (50, 50, 50), will change results of our analysis. Such large values can only be used if such belief is supported by prior biological knowledge.

**Graphical summary of proposed model-based clustering method**

Our model-based clustering method can be regarded as a mixture of Bayesian hierarchical models. Figure 2 shows the directed acyclic graphic of our mixture of Bayesian hierarchical models. The shaded areas on the left and right sides of the figure contain information in cases and controls respectively. Cases and controls are linked by the shared information displayed in the center part of the figure.

**Inferences, the decision rule for calling significant SNPs**

Calling which SNPs are significantly associated with the outcome is equivalent to assigning SNPs to cluster + or cluster −. The decision is made based on the posterior probability of a cluster[51].

Conditional on all observed data and hyper-parameters in the prior, we derive the posterior probability of cluster membership (also called "responsibility" in the machine learning community) using Bayesian theorem as

$$\gamma_{g,k} = \Pr(z_{g,k} = 1 | S_g, \pi, \alpha, \beta) = \frac{\pi_k \xi_k(S_g | \alpha, \beta)}{\sum_{j \in \{0,+,-\}} \pi_j \xi_j(S_g | \alpha, \beta)} \qquad (2)$$

where $\xi_k(S_g | \alpha, \beta)$ is the marginal density of genotypes of the g-th SNP in the k-th cluster of all patients. (derivation on the marginal density $\xi_k$ is given in Section A of Supplementary File S3).

To estimate the responsibilities $\gamma_{g,k}$, we first apply the EM algorithm to estimate the model parameters, mixture weights $\pi_k$, and then plug the estimated model parameters into Formula (2). See Section B of Supplementary File S3 for details on the implementation.

A straightforward decision rule about cluster membership is to assign each SNP to the cluster with the highest posterior probability, i.e., assign SNP g to the cluster corresponding to the largest value of $\gamma_{g,0}$, $\gamma_{g,+}$, and $\gamma_{g,-}$.

An alternative is to assign a SNP to effective clusters (+ or −) if its responsibility of coming from cluster + or − is greater than a threshold $\tau$. Following Yuan and Kendziorski (2006)[52], the value of $\tau$ can be specified to achieve a desired level of false discovery rate (FDR) given as follows[53]:

$$\widehat{FDR} = \frac{\Sigma_{\{g:\hat{\gamma}_{g,+}>\tau,\text{or }\hat{\gamma}_{g,-}>\tau\}} \hat{\gamma}_{g,0}}{\text{card}\{g:\hat{\gamma}_{g,+}>\tau,\text{or }\hat{\gamma}_{g,-}>\tau\}} \qquad (3)$$

where "card{set}" means the number of elements in "set".

The second approach is preferred in most applications, since controlling FDR of detected SNPs is often desired for genomic studies. Users can decide which decision rule to use based on their specific applications.

**Design of simulation studies**

We conducted simulation studies to compare the performance of our model-based clustering method with the SNP-wise approach (e.g., logistic regression followed by multiple testing adjustment). We generated datasets using different settings, by varying the combination of factors, including sample sizes (total 200 or 1000 with half cases and half controls), number of SNPs (1000, 20000, 500000), and various mixture proportions $\pi$. Details of multiple settings of simulation studies are given in Supplementary Table S1.

In addition, to investigate the robustness of our method against the misspecification of the MAF prior, we used four settings of truncated beta distribution $\text{Beta}(\alpha, \beta)$ with the range [0.05, 0.5] for data generation. The four settings included truncated $\text{Beta}(2, 5)$, which was also used in data analysis; truncated $\text{Beta}(2, 4)$ and $\text{Beta}(1.5, 3.5)$ distributions with a shifted mode to the right and left-hand side compared to $\text{Beta}(2, 5)$ respectively; and truncated $\text{Beta}(1.5, 5.5)$ with a sharper peak than $\text{Beta}(2, 5)$. The last 3 settings were used to investigate the performance of our method when MAF priors were incorrectly specified. All these four prior distributions were truncated to the range $(0.05, 0.5)$, which ensured the MAFs of simulated SNPs were always greater than 0.05 and smaller than 0.5.

For each combination of settings above, we simulated 100 datasets. For every simulated dataset, we analyzed it using three approaches: (1) the traditional SNP-wise approach; (2) our method with prior of MAFs set as the Beta-approximation of truncated $\text{Beta}(2, 5)$; and (3) our method with prior of MAFs set as the Beta-approximation of empirical MAF distribution estimated from simulated SNP data (see Section C of Supplementary File S3).

In data analysis, some initial values need to be specified to conduct the EM algorithm. We used raw p-values from the SNP-wise approach to specify initial cluster membership (SNPs with raw P-value smaller than 0.05 and different estimated MAF in cases/controls were initially classified into cluster $+/-$, while other SNPs were initially classified into cluster 0), and then we calculated initial values of $\boldsymbol{\pi}$ based on initial cluster memberships. Dirichlet distribution with concentration parameter 3 (i.e., $\text{Dirichlet}(3,3,3)$) was used as the (weak) prior distribution for $\boldsymbol{\pi}$.

**Comparison criteria in simulation studies**

Two criteria were used to compare our method with the SNP-wise approach in the simulation studies: FDR and sensitivity. Unlike real data analysis, actual FDR of analyses can be calculated in the simulation studies, since the truth of effective SNPs is known in these studies. We calculated actual FDR as the proportion of truly non-effective SNPs among all SNPs called as significant by data analysis. Both our method and the traditional approach targeted FDR to be controlled at a level of 0.05, thus the successful method should have the actual FDR closer to 0.05. Sensitivity is defined as the proportion of SNPs detected significant among truly effective SNPs. Higher sensitivity means the method is more powerful.

Note that specificity is usually evaluated together with sensitivity. We did not report specificity in this article since both FDR and 1−specificity are measures of rate of type I error (false positive rate). FDR is much more popularly used in genomic studies. Hence, we decided to control FDR instead of specificity.

## Data availability

The two GWAS datasets are downloaded from the Gene Expression Omnibus (GEO) (https://www.ncbi.nlm.nih.gov/geo/) with accession IDs GSE65777 and GSE66903. We are wrapping our codes into an R package, called "BayesGWAS", and will submit to Bioconductor soon.

## References

1.  Wu, M.C., et al., *Powerful SNP-set analysis for case-control genome-wide association studies.* Am J Hum Genet, 2010. **86**(6): p. 929-42.
2.  Dai, H., et al., *Weighted SNP set analysis in genome-wide association study.* PLoS One, 2013. **8**(9): p. e75897.
3.  Lu, Z.H., et al., *Multiple SNP Set Analysis for Genome-Wide Association Studies Through Bayesian Latent Variable Selection.* Genet Epidemiol, 2015. **39**(8): p. 664-77.
4.  Cologne, J., et al., *Stepwise approach to SNP-set analysis illustrated with the Metabochip and colorectal cancer in Japanese Americans of the Multiethnic Cohort.* BMC Genomics, 2018. **19**(1): p. 524.
5.  Fridley, B.L. and J.M. Biernacka, *Gene set analysis of SNP data: benefits, challenges, and future directions.* Eur J Hum Genet, 2011. **19**(8): p. 837-43.
6.  Kang, H.M., et al., *Variance component model to account for sample structure in genome-wide association studies.* Nat Genet, 2010. **42**(4): p. 348-54.
7.  Lippert, C., et al., *FaST linear mixed models for genome-wide association studies.* Nat Methods, 2011. **8**(10): p. 833-5.
8.  Zhou, X. and M. Stephens, *Genome-wide efficient mixed-model analysis for association studies.* Nat Genet, 2012. **44**(7): p. 821-4.
9.  Goddard, M.E., et al., *Genetics of complex traits: prediction of phenotype, identification of causal polymorphisms and genetic architecture.* Proc Biol Sci, 2016. **283**(1835).
10. Chen, C., J.P. Steibel, and R.J. Tempelman, *Genome-Wide Association Analyses Based on Broadly Different Specifications for Prior Distributions, Genomic Windows, and Estimation Methods.* Genetics, 2017. **206**(4): p. 1791-1806.
11. Mallick, H. and N. Yi, *Hierarchical Models for Genetic Association Studies.* Journal of Biometrics and Biostatistics, 2013. **4**: p. e124.
12. Fernando, R.L. and D. Garrick, *Bayesian methods applied to GWAS.* Methods Mol Biol, 2013. **1019**: p. 237-74.

13. Wang, Q., et al., *An efficient empirical Bayes method for genomewide association studies.* J Anim Breed Genet, 2016. **133**(4): p. 253-63.
14. Sanyal, N., et al., *GWASinlps: non-local prior based iterative SNP selection tool for genome-wide association studies.* Bioinformatics, 2019. **35**(1): p. 1-11.
15. Newton, M.A., et al., *On differential variability of expression ratios: improving statistical inference about gene expression changes from microarray data.* J Comput Biol, 2001. **8**(1): p. 37-52.
16. Kendziorski, C.M., et al., *On parametric empirical Bayes methods for comparing multiple groups using replicated gene expression profiles.* Stat Med, 2003. **22**(24): p. 3899-914.
17. Lo, K. and R. Gottardo, *Flexible empirical Bayes models for differential gene expression.* Bioinformatics, 2007. **23**(3): p. 328-35.
18. Li, Y., et al., *Detecting disease-associated genomic outcomes using constrained mixture of Bayesian hierarchical models for paired data.* PLoS One, 2017. **12**(3): p. e0174602.
19. Qiu, W., et al., *A marginal mixture model for selecting differentially expressed genes across two types of tissue samples.* Int J Biostat, 2008. **4**(1): p. Article 20.
20. Robinson, M.D. and G.K. Smyth, *Moderated statistical tests for assessing differences in tag abundance.* Bioinformatics, 2007. **23**(21): p. 2881-7.
21. McCarthy, D.J., Y. Chen, and G.K. Smyth, *Differential expression analysis of multifactor RNA-Seq experiments with respect to biological variation.* Nucleic Acids Res, 2012. **40**(10): p. 4288-97.
22. Anders, S. and W. Huber, *Differential expression analysis for sequence count data.* Genome Biol, 2010. **11**(10): p. R106.
23. Love, M.I., W. Huber, and S. Anders, *Moderated estimation of fold change and dispersion for RNA-seq data with DESeq2.* Genome Biol, 2014. **15**(12): p. 550.
24. Raab, M.S., et al., *Multiple myeloma.* Lancet, 2009. **374**(9686): p. 324-39.
25. Adams, J., *The development of proteasome inhibitors as anticancer drugs.* Cancer Cell, 2004. **5**(5): p. 417-21.
26. Altun, M., et al., *Effects of PS-341 on the activity and composition of proteasomes in multiple myeloma cells.* Cancer Res, 2005. **65**(17): p. 7896-901.
27. Field-Smith, A., G.J. Morgan, and F.E. Davies, *Bortezomib (Velcadetrade mark) in the Treatment of Multiple Myeloma.* Ther Clin Risk Manag, 2006. **2**(3): p. 271-9.
28. Broyl, A., et al., *Mechanisms of peripheral neuropathy associated with bortezomib and vincristine in patients with newly diagnosed multiple myeloma: a prospective analysis of data from the HOVON-65/GMMG-HD4 trial.* Lancet Oncol, 2010. **11**(11): p. 1057-65.
29. Magrangeas, F., et al., *A Genome-Wide Association Study Identifies a Novel Locus for Bortezomib-Induced Peripheral Neuropathy in European Patients with Multiple Myeloma.* Clin Cancer Res, 2016. **22**(17): p. 4350-4355.
30. Schiff, D., P.Y. Wen, and M.J. van den Bent, *Neurological adverse effects caused by cytotoxic and targeted therapies.* Nat Rev Clin Oncol, 2009. **6**(10): p. 596-603.
31. Richardson, P.G., et al., *Proteasome inhibition in hematologic malignancies.* Ann Med, 2004. **36**(4): p. 304-14.
32. Dimopoulos, M.A., et al., *Risk factors for, and reversibility of, peripheral neuropathy associated with bortezomib-melphalan-prednisone in newly diagnosed patients with* *multiple myeloma: subanalysis of the phase 3 VISTA study.* Eur J Haematol, 2011. **86**(1): p. 23-31.
33. Beijers, A.J., J.L. Jongen, and G. Vreugdenhil, *Chemotherapy-induced neurotoxicity: the value of neuroprotective strategies.* Neth J Med, 2012. **70**(1): p. 18-25.
34. Lanzani, F., et al., *Role of a pre-existing neuropathy on the course of bortezomib-induced peripheral neurotoxicity.* J Peripher Nerv Syst, 2008. **13**(4): p. 267-74.
35. Bruna, J., et al., *Evaluation of pre-existing neuropathy and bortezomib retreatment as risk factors to develop severe neuropathy in a mouse model.* J Peripher Nerv Syst, 2011. **16**(3): p. 199-212.
36. Meregalli, C., *An Overview of Bortezomib-Induced Neurotoxicity.* Toxics, 2015. **3**(3): p. 294-303.
37. Johnson, D.C., et al., *Genetic factors underlying the risk of thalidomide-related neuropathy in patients with multiple myeloma.* J Clin Oncol, 2011. **29**(7): p. 797-804.
38. Corthals, S.L., et al., *Genetic factors underlying the risk of bortezomib induced peripheral neuropathy in multiple myeloma patients.* Haematologica, 2011. **96**(11): p. 1728-32.
39. Favis, R., et al., *Genetic variation associated with bortezomib-induced peripheral neuropathy.* Pharmacogenet Genomics, 2011. **21**(3): p. 121-9.
40. Welter, D., et al., *The NHGRI GWAS Catalog, a curated resource of SNP-trait associations.* Nucleic Acids Res, 2014. **42**(Database issue): p. D1001-6.
41. MacArthur, J., et al., *The new NHGRI-EBI Catalog of published genome-wide association studies (GWAS Catalog).* Nucleic Acids Res, 2017. **45**(D1): p. D896-D901.
42. Campo, C., et al., *Bortezomib-induced peripheral neuropathy: A genome-wide association study on multiple myeloma patients.* Hematol Oncol, 2018. **36**(1): p. 232-237.
43. Manolio, T.A., et al., *Finding the missing heritability of complex diseases.* Nature, 2009. **461**(7265): p. 747-53.
44. Johansson, C., et al., *Structural analysis of human KDM5B guides histone demethylase inhibitor development.* Nat Chem Biol, 2016. **12**(7): p. 539-45.
45. Tumber, A., et al., *Potent and Selective KDM5 Inhibitor Stops Cellular Demethylation of H3K4me3 at Transcription Start Sites and Proliferation of MM1S Myeloma Cells.* Cell Chem Biol, 2017. **24**(3): p. 371-380.
46. Park, T. and G. Casella, *The Bayesian Lasso.* Journal of the American Statistical Association, 2008. **103**(482): p. 681-686.
47. Li, Q. and N. Lin, *The Bayesian elastic net.* Bayesian Analysis, 2010. **5**(1): p. 151-170.
48. Hazimeh, H. and C. Zhai, *Axiomatic Analysis of Smoothing Methods in Language Models for Pseudo-Relevance Feedback*, in *International Conference on The Theory of Information Retrieval*. 2015, ACM, New York, NY , USA: Northampton, Massachusetts, USA. p. 141-150.
49. Valcarce, D., J. Parapar, and Á. Barreiro, *Additive Smoothing for Relevance-Based Language Modelling of Recommender Systems*, in *the 4th Spanish Conference on Information Retrieval*. 2016, ACM   New York, NY , USA: Granada, Spain. p. Article No. 9.
50. Keinan, A., et al., *Measurement of the human allele frequency spectrum demonstrates greater genetic drift in East Asians than in Europeans.* Nat Genet, 2007. **39**(10): p. 1251-5.

51. Pan, W., J. Lin, and C.T. Le, *Model-based cluster analysis of microarray gene-expression data.* Genome Biol, 2002. **3**(2): p. RESEARCH0009.
52. Yuan, M. and C. Kendziorski, *A unified approach for simultaneous gene clustering and differential expression identification.* Biometrics, 2006. **62**(4): p. 1089-98.
53. Newton, M.A., et al., *Detecting differential gene expression with a semiparametric hierarchical mixture method.* Biostatistics, 2004. **5**(2): p. 155-76.

## Acknowledgements

Thanks Dr. Stéphane Minvielle for helpful discussion about QC steps in their paper [8].

Thanks Dr. Leland Wilkinson for helpful discussion and comments about paper revision at "2018 NISS Writing Workshop for Junior Researchers in Statistics and Data Science".

## Author contributions statement

WQ, XZ, LX, and JS conceived and designed the study. XZ, LX, and YX performed the data analysis and wrote the paper. WQ and JS commented and revised the paper. All authors read and approved the final manuscript.

## Competing interests

The authors declare that they have no competing interests.

## Tables

**Table 1. Significant SNPs detected by our method based on the discovery dataset (GSE65777).**

| Pseudo count | $\widehat{FDR}$ | Detected SNPs |
|---|---|---|
|  | $< 0.1$ | rs10862339  rs1344016  rs2839629* |
| (3,3,3) | $< 0.05$ | rs10862339  rs1344016 |
|  |  | rs10862339  rs1344016  rs2414277 |
|  | < 0.1 | rs2839629*  rs4351714**  rs4776196 |
| (20,20,20) | < 0.05 | rs10862339  rs1344016 |

The SNP labeled with '*' is the only SNP reported by Magrangeas et al. (2016)[29] as validated SNP, the SNP labeled with '**' is a novel SNP detected by our method, and all other SNPs in the table are reported by Magrangeas et al. (2016)[29] in discovery data, but not validated in replication data.

**Table 2. Validation results for SNPs listed in Table 1. One-sided logistic regression with permutations followed by FDR adjustment for the validation set (GSE66903).**

| SNP_ID | Odds ratio | P.raw | P.permutation | P.FDR |
|---|---|---|---|---|
| rs10862339 | 0.98 (0.57-1.69) | 0.4662 | 0.4783 | 0.4783 |
| rs1344016 | 1.05 (0.59-1.86) | 0.4329 | 0.425 | 0.4783 |
| rs2839629 | 2.02 (1.12-3.65) | 0.0096 | 0.0108 | 0.0324 |

P.raw: raw P-values from one-sided logistic regression; P.permutation: P-values determined by permutation; P.FDR: permuted P-values after FDR adjustment.

**Figure legends**

**Figure 1. Simulation results for 500,000 SNPs with 200 effective SNPs using four different MAF distributions for data generation and two different sample sizes respectively.** The upper and lower four rows contain results of simulated data with 200 samples and 1,000 samples respectively. The left panel shows truncated MAF distributions for data generation (solid line) and prior distributions for analysis

(approximated beta distributions via the moment matching: dashed lines are Beta-approximations of truncated $\text{Beta}(2,5)$ and dotted lines are Beta-approximations of empirical distributions estimated from data). The middle panel shows boxplots for FDR and the vertical dashed line represents the nominal level (0.05). The right panel shows boxplots for paired difference of sensitivity between our method (using truncated $\text{Beta}(2,5)$ or empirical distribution for analysis) and the SNP-wise approach, and the vertical dashed line represents 0 (i.e., same performance between our method and the SNP-wise approach). White boxes represent the SNP-wise approach, light grey boxes represent our method using truncated $\text{Beta}(2,5)$ for analysis, and dark grey boxes represent our method using empirical distributions for analysis.

**Figure 2. Directed acyclic graph representation of our model-based clustering method.** Observed data (SNP genotypes), cluster memberships, MAFs, and mixture proportions are denoted by $S$, $z$, $\theta$, and $\pi$ respectively. Plain solid rectangles represent observations. Diamonds represent latent variables of unknown cluster membership. Plain solid circles indicate model parameters to be estimated, while dashed circles represent nuisance parameters to be integrated out from the model likelihood by marginalization. Gray-filled circles represent pre-specified hyper-parameters.

## Supplementary information

**Supplementary Table S1**: **Different settings to generate data for both 200-sample and 1000-sample simulations.** In data analysis, for prior MAF distribution, we used both truncated $\text{Beta}(2,5)$ and empirical distribution estimated from MAFs of all SNPs in each simulated data. Note all these prior MAF distributions used for data analysis were approximated by a Beta distribution via the moment matching. Truncated $\text{Beta}(2,5)$ was approximated by $\text{Beta}(3.29, 9.56)$.

**Supplementary Table S2. P-values from Wilcoxon signed rank tests between our method and the SNP-wise approach on sensitivity and |FDR-0.05| for all settings of our simulation studies.**

**Supplementary File S3. Supplementary materials and methods**. Details for marginalization, EM implementation, moment matching, and QC for real data analysis.

**Supplementary File S4. Simulation results for different SNP sizes with different numbers of effective SNPs using 4 different MAF distributions for data generation and 2 different sample sizes respectively.** Includes Figures S1-S5. In each figure, the upper and lower 4 rows contain results of simulated data with 200 samples and 1,000 samples respectively. The left panel shows truncated MAF distributions for data generation (solid line) and prior distributions for analysis (approximated beta distributions via the moment matching: dashed lines are Beta-approximations of truncated $\text{Beta}(2,5)$ and dotted lines are Beta approximations of empirical distributions estimated from data). The middle panel shows boxplots for FDR and the vertical dashed line represents the nominal level (0.05). The right panel shows boxplots for paired difference of sensitivity between our method (using truncated $\text{Beta}(2,5)$ or empirical distribution for analysis) and the SNP-wise approach, and the vertical dashed line represents 0 (i.e., same performance between our method and the SNP-wise approach). White boxes represent the SNP-wise approach, light grey boxes represent

our method using truncated $\text{Beta}(2,5)$ for analysis, and dark grey boxes represent our method using empirical distributions for analysis.

**Supplementary File S5. Additional information on datasets GSE66903 and GSE65777.** Two subtables are included. S5 Table (a) contains the top 1000 ranked SNPs in the discovery set with smallest raw P-values and S5 Table (b) contains the 30 SNPs that also have a raw P-value <0.05 for the validation set.